\documentclass[sigconf]{acmart}

\usepackage{booktabs} % For formal tables

\usepackage{algorithm}
\usepackage{algorithmic}
\usepackage{color}
\usepackage{xspace}

\setcopyright{rightsretained}
\acmDOI{10.475/123_4}

\acmISBN{123-4567-24-567/08/06}

\acmConference[KDD'17]{ACM KDD conference}{August 2017}{Halifax, Nova Scotia Canada} 
\acmYear{2017}
\copyrightyear{2017}

\acmPrice{15.00}

\newcommand\note[2]{{\color{#1}\bf #2}}
\newcommand\sofa[1]{{\note{blue}{SA: #1}}}
\newcommand\kha{\texttt{Kharita}\xspace}
\newcommand\khastar{\texttt{Kharita$^*$}\xspace}

\begin{document}
%\title{Incremental Map Inference from GPS Data}
\title{\kha: Robust Map Inference using Graph Spanners}
%\titlenote{Produces the permission block, and copyright information}
%\subtitle{Extended Abstract}
%\subtitlenote{The full version of the author's guide is available as \texttt{acmart.pdf} document}

\author{Rade Stanojevic$^1$, Sofiane Abbar$^1$, Saravanan Thirumuruganathan$^1$, Sanjay Chawla$^1$ \and Fethi Filali$^2$, Ahid Aleimat$^2$}
%\authornote{}
\affiliation{%
  \institution{$^1$Qatat Computing Research Institute, HBKU\\
  P.O. Box 5825, Doha\\
  $^2$Qatar Mobility Innovation Center, QSTP\\
  P.O. Box 210531, Doha}
  %\streetaddress{P.O. Box 1212}
  \state{Qatar} 
}
\email{{rstanojevic, sabbar, sthirumuruganathan,schawla}@hbku.edu.qa, {filali,ahide}@qmic.com}

\renewcommand{\shortauthors}{R. Stanojevic et al.}

\begin{abstract}
    The widespread availability of GPS information in everyday devices such as cars, smartphones and smart watches make it possible to collect large amount of geospatial trajectory information. A particularly important, yet technically challenging, application of this data is to identify the underlying road network and keep it updated under various changes. In this paper, we propose efficient algorithms that can generate accurate maps in both batch and online settings. Our algorithms utilize techniques from graph spanners so that they produce maps can effectively handle a wide variety of road and intersection shapes. We conduct a rigorous evaluation of our algorithms over two real-world datasets and under a wide variety of performance metrics. Our experiments show a significant improvement over prior work. In particular, we observe an increase in Biagioni $f$-score of up to 20\% when compared to the state of the art while reducing the execution time by an order of magnitude. We also make our source code open source for reproducibility and enable other researchers to build on our work.
\end{abstract}

% We no longer use \terms command


\maketitle

\section{Introduction}
\label{sec:introduction}

\begin{comment}
\sofa{Motivate the timely relevance of map inference. In July 2016 the Financial Times reported that Uber is set to invest \$500 million into a global mapping project \footnote{https://www.ft.com/content/e0dfa45e-5522-11e6-befd-2fc0c26b3c60}. Toyota has also presented in CES 2016 an ambitious project for creating maps using GPS devices \footnote{https://www.cnet.com/au/news/toyota-wants-to-make-google-mapping-tech-obsolete/}}
\end{comment}

\noindent {\bf Map Construction from Crowd-sourced GPS Data:} The widespread adoption of smartphones with GPS and sensors like accelerometers makes it possible to collect large amounts of geo-spatial trajectory data. This has given rise to an important data science research question: \emph{Is it possible to accurately infer the underlying road network solely from GPS trajectories?} An affirmative answer to this question will have a profound impact on map making. It will reduce the cost and democratize map construction and reduce the lag between changes in the underlying road network (e.g., due to construction) and its reflection in the map.  The latter is particularly useful if autonomous vehicles are to become part of the mainstream traffic profile. For instance, Uber has announced last July that it will invest \$500 million into a global mapping project\footnote{https://www.ft.com/content/e0dfa45e-5522-11e6-befd-2fc0c26b3c60} and Toyota has showcased in CES 2016 its ambitious project for creating maps using GPS devices\footnote{https://www.cnet.com/au/news/toyota-wants-to-make-google-mapping-tech-obsolete/}.  Not surprisingly, there has been extensive prior work by the data mining community \cite{ChenLHYGG16,LiuBEWFZ12,schroedl2004mining,BiagioniE12} for map creation.  However, most of them exhibit a large gap in quality between inferred and actual road maps. 

\noindent {\bf Challenges for Accurate Map Inference:} There are a number of technical challenges that a map construction algorithm using crowdsourced data needs to overcome including: (i) GPS sensors in smartphones, while reliable in general, can have non-trivial {\em GPS errors}.  The errors are often acute in ``urban canyons'' with dense buildings or other structures; (ii) There is a substantial  {\em disparity in data} due to the opportunistic data collection from smartphones. For example, while popular highways might have large number of data points, residential areas might only have handful of points; (iii) There exist a wide variety of {\em sampling rates} in which the GPS information is collected, often due to power concerns; (iv) Existing algorithms tend to overfit on the data sets on which they were tested. For example, many  existing algorithms make some {\em implicit assumptions} that are specific to the road structures common to the United States and  Europe. For instance, many countries in the  Middle East have circular intersections (roundabouts) that  are notoriously hard for existing algorithms to capture; (v) Finally, real-world roads often undergo various changes such as closure, new construction, shifting, temporary blocking due to accidents etc. Despite this, most prior work often treat the map construction as a {\em static problem}, resulting in maps that eventually become outdated. For example, in the city of Doha, with a dynamic road infrastructure, both Google maps and OpenStreetMap take over a year to update several major changes to the road network causing frustration to many drivers.

%These various challenges often results in a substantial gap in quality between the maps produced by previous map construction algorithms and maps obtained from road surveys.
\begin{figure}[t]
    \includegraphics[width=0.5\textwidth]{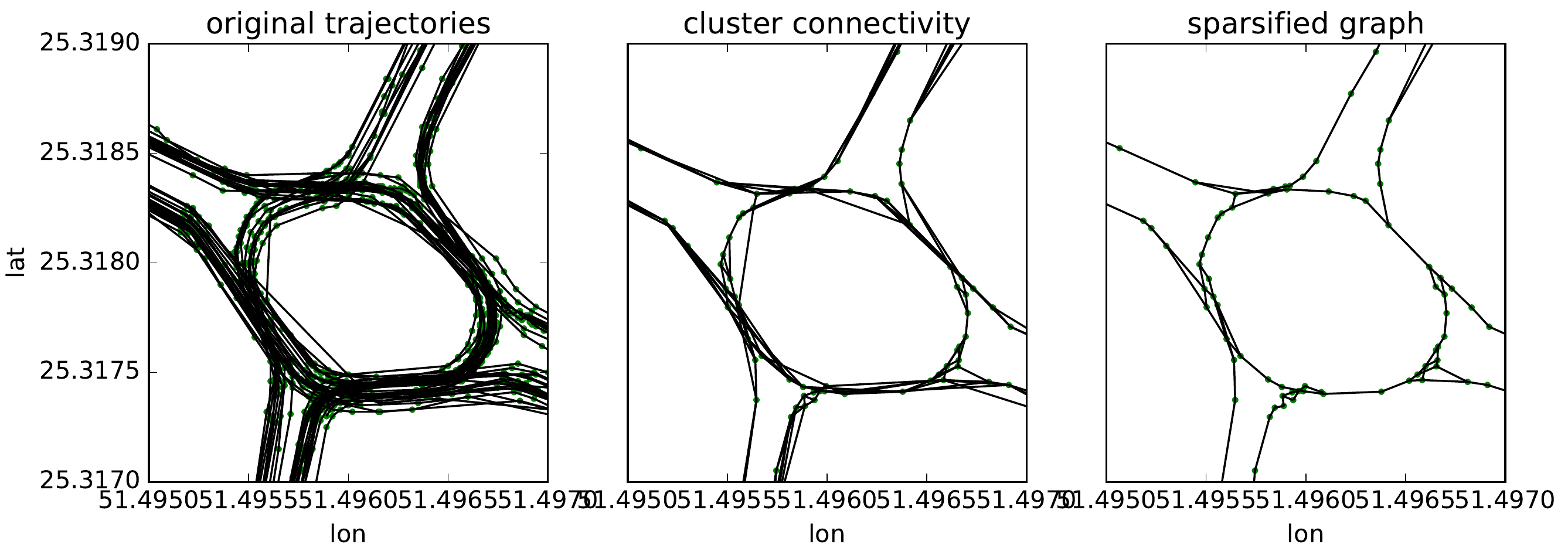}
    \caption{{\bf \kha process:} All raw trajectories passing through a roundabout (left). Centroids of the clusters and graph $G=(V,E_C)$ obtained from the raw trajectories projected onto clusters (middle). Final Kharita map obtained after sparsifying $G$ (using a spanner).}
    \label{fig:offlineAlgoSteps}
\end{figure}

\noindent {\bf Crossing the Quality Chasm:} In this paper, we propose concrete steps towards reducing the gap between the \emph{inferred} and the \emph{true} map  by deploying a diverse collection of tools and techniques: 

\begin{comment}
  \item {\bf Leveraging Angle and Speed Information:} 
        Modern GPS navigators and smartphones are often capable of providing additional accurate information such as speed and angle
        in addition to standard GPS information such as latitude, longitude and timestamp.
        However, most of the existing work on automatic map inferrence does not utilize this additional information resulting in reduced map quality.
        In our paper, we extensively use this information in myriad ways resulting in a better map. 
\end{comment}

{\bf Principled Approach through Graph Spanners:}
We propose \kha\footnote{Kharita means map in Arabic.} -- a two-phase approach -- that combines clustering with a map construction algorithm based on graph spanners.  Graph spanners are subgraphs that 
approximate the shortest path distance of the original graph.  We show that these provide us a parameterized way to obtain sparse maps that retain the connectivity information effectively.
The spanner formulation is particularly useful in handling different types of  road and intersection shapes. 

{\bf Online Map Construction:} We propose \khastar an online version of the 
map construction algorithm which combines the clustering and graph spanner phases. 
    %\item {\bf Incremental Update of Maps:}
    %    Another major deficiency of prior works is that they often treat the map construction as a static problem.
    %    We investigate and evaluate two complementary approaches for producing up-to-date maps.

{\bf Rigorous Evaluation:}
We conduct extensive evaluation of our approaches using standard performance metrics that 
include comparing the true and the inferred maps using geometric and topological metrics.
        %We also propose a series of metrics that could be used to evaluate incremental map algorithms.
        %Based on these evaluations, we highlight a number of future opportunities for improvement and research.

{\bf Reproducibility:} 
In addition to providing high level details about our algorithms, we also make our source code open  
so that other researchers can readily build on top our work. The code can be found at this url: \url{https://github.com/vipyoung/kharita} 
        %Additionally, we dedicate an entire section (Section~\ref{sec:tipsAndTricks}) for providing some tips, tricks and other useful rule of thumb.

\noindent {\bf Solution in a nutshell:}

We briefly describe our offline solution. We abstract the map inference task as that
of constructing a directed graph $G=(V,E)$ from a collection of GPS trajectories. 
We proceed as follows:
\begin{enumerate}
\item We begin by first treating trajectories as a collection of points, by ignoring the implicit ordering in the trajectories.
\item The next step is clustering these points into those that belong to the same road segments using the location and heading information. 
\item We then bring back the trajectory information and assign each point in the trajectory
to its cluster center. If two consecutive points in the trajectory $tr$, say $tr:x_{i}$
and $tr:x_{i+1}$ are assigned to cluster centers $u$ and $v$ respectively,
then we create an edge from $u$ to $v$. We now have a graph $G=(V,E_C)$ whose nodes are cluster centers derived in the previous step.
\item In the final step, the edge set $E_C$ is pruned to $E$ by invoking a spanner algorithm which
sparsifies the graph while maintaining the connectivity and approximate shortest path distances. 
\end{enumerate}

Figure~\ref{fig:offlineAlgoSteps} provides an illustration of the major steps of the offline algorithm. While our approach is conceptually simple it significantly outperforms the state-of-the-art by improving the Biagioni TOPO (the de-facto standard metric for measuring map quality, see Sec. \ref{sec:evaluation}) f-score of up to 20\%. The uplift in performance comes mainly from the efficient use of angle and speed information and elegant handling of geographic information (via clustering) and topological structure (via spanner). 

%The rest of the paper is structured as follows. In Section~\ref{sec:dataModel}, we introduce the notation and formally define the problem. Section~\ref{sec:offalgo} introduces \kha, our two-step offline algorithm while Section~\ref{sec:onlineAlgorithm} introduces the corresponding online variant \khastar. Experimental details can be found in Section~\ref{sec:evaluation}. We discuss the related work in Section~\ref{sec:relatedWork} and conclude in Section~\ref{sec:conclusion} with a summary and directions for future work.

\section{Definitions and Problem Statement}
\label{sec:dataModel}

\begin{definition}
A GPS point is a five-tuple $(lat,lon,t,s,a)$, where $lat$ is the latitude, $lon$ the longitude,
$t$ the timestamp, $s$ the speed and $a$ the angle.
\end{definition}
\begin{definition}
A trajectory $tr$ is a chronologically ordered collection of GPS points. We represent
a trajectory $tr$ as $\{x_1, x_2, \ldots, x_{|tr|}\}$ where each $x_i$ is a GPS point and
$|tr|$ is number of GPS measurements in trajectory $tr$.
\end{definition}
\begin{definition}
A geometric, weighted and directed  graph $G=(V,E)$ consists of vertex set $V$ where each $v \in V$ is a $(lat,lon, angle)$ triplet.
\end{definition}
\begin{definition}
The length of the shortest path between two nodes $u$ and $v$ in $G$ is denoted as $d_{G}(u,v)$.
\end{definition}
\begin{definition}
A graph $H =(V,E')$ is an $(\alpha > 1)$ spanner of $G =(V,E)$ if $E' \subset E$ and for any pair of nodes $u$ and $v$, $d_{G}(u,v) \leq d_{H}(u,v) \leq \alpha*d_{G}(u,v)$. 
\end{definition}

\subsection{Problem Definition}
\noindent
{\bf Given:} A  collection $T$ of GPS trajectories. \\[2ex]
{\bf Objective:} Infer a geometric, directed and weighted $G$ which
represents the underlying road network. \\

\begin{comment}
\section{Data Model(0.5 page)}
\label{sec:mapInferencePipeline}
We collect GPS data points opportunistically from smartphones and GPS navigators.
Each GPS data point is a quintuple of {\em latitude, longitude, timestamp, speed} and {\em direction}.
\textcolor{red}{S: define how angle is computed/how to interpret it}.
Often, the GPS data points are subject to noise that can be anywhere 
from 10 meters to more than 100 meters especially in urban canyons where tall structures obstruct GPS measurements.
A GPS trajectory is a set of chronologically ordered GPS points from the same vehicle.
We represent a trajectory $t$ as $\{x_1, x_2, \ldots, x_{|t|}\}$ where each $x_i$ is a GPS measurement and 
$|t|$ is number of GPS measurements in trajectory $t$.
We denote the set of all trajectories available for map inference as $T = \{t_1, t_2, \ldots, t_n\}$.
The $j^{th}$ GPS measurement of trajectory $t_i$ is denoted as $t_i[j]$.

Our objective is to process the trajectory collection to produce a routable map.
We represent the map as a directed graph $G=(V,E)$ where 
each vertex $v \in V$ corresponds to a geo-location 
while a directed edge connects nodes $(u,v)$ if there exists a road segment between the corresponding geo-locations.
Most of the prior work represent the map using this approach.

Using the definitions above, the map construction problem can be specified as:
Given a GPS trajectory collection $T$, construct a routable directed graph $G=(V,E)$ that best summarizes $T$. 

Performance metrics
Sanjay's problem statement.
\end{comment}

\begin{table}
\small
\caption{Notation}
\label{tbl:notation}
\centering
\begin{tabular}{c|l}\hline\hline
$x_i$ & GPS point: triplet (longitude, latitude, angle) \\
$tr$ &  Trajectory: sequence of GPS points $\{x_1, \ldots x_{|tr|}\}$ \\
$|tr|$ &  Number of GPS points in $tr$ \\
$G$ & Road networks:  directed graph $G=(V, E)$ \\
$V$ & Vertices of $G$: each vertex is a cluster centroid \\
$E$ & Edges of $G$: each edge $e(.,.)$ is a road segment \\
$H$ & Spanner graph of $G$ \\
$\alpha$ & Spanner stretch of $H$ \\
\hline
$v(x_i,x_j)$ & Vincenty geometric distance in meters  \\
$d_{G}(u,v)$ & Length in meters of the shortest path \\
& between $u$ and $v$ in $G$ \\
$d_{\theta}(x_i, x_j)$ & Distance between two gps points that combines\\
& their locations and angles\\ 
\hline
$cr$ & Clustering radius (\kha, \khastar)\\
$sr$ & Densification rate (\kha, \khastar) \\
$\theta$ & Penalty of the angle difference (\kha) \\
$ha$ & Tolerance of angle difference (\khastar) \\
\end{tabular}
\end{table}

\section{\kha: Offline Algorithm}
\label{sec:offalgo}
In this section, we describe \kha, a two-phase algorithm that processes a collection of GPS trajectories to produce a routable map.
%\sofa{@Rade uses a gps point notation (L, alpha) different from what we have in the definitions }
\begin{comment}
\subsection{Data}
Data we use to generate the map is crowdsourced by by a fleet of several hundred vehicles with attached GPS-enabled blackbox devices. These devices report a number of variables about the state of the car via a cellular network. The data is reported using timestamped records which are generated every 10 seconds (or less) including the following information:
the \textit{location} $(lon,lat)$ of the vehicle, the \textit{speed} and the \textit{heading} of the moving direction. Heading is measured in angles against the North axis in degrees reporting values from $0$  to $360^{\circ}$. Note that most of prior work on map inference \cite{} does not utilize the heading information even though it is standardly reported by GPS-enabled devices. 

The data used in this paper covers a rectangle (in $lat,lon$ coordinates) approximately 6km$\times$8km of urban region in the city of Doha with a mixture of highways, high and medium volume roads and capillary streets. Our data contains all the records generated by our fleet within this rectangle in October and November of 2015. We have overall around 12M datapoints, or 200K records per day. 

\textbf{RS: do we want to discuss privacy/annonimity issues of the data?}
\end{comment}

\subsection{Distance Metric}
\label{subsec:distanceMetric}
A key novelty of our paper is the extensive use of the {\tt angle} information that is widely available as part of GPS output.
One can measure the distance between two (latitude, longitude) pairs ($L_1, L_2$) using Vincenty distance formula.
We denote it by $v(L_1, L_2)$ and it provides the distance in meters.
The distance between two angles can be computed using the unit circle metric defined as
$d_{\circ}(\alpha_1,\alpha_2) = min(|\alpha_1-\alpha_2|,360^\circ-|\alpha_1-\alpha_2|)$.
For example, the distance between $350^{\circ}$ and $10^{\circ}$ is $20^{\circ}$.
In order to compute the distance between two GPS points, we should consider both the location and the heading (angle of movement).
Given two GPS points $(L_1,\alpha_1)$ and  $(L_2,\alpha_2)$, we combine the aforementioned metrics as

\begin{equation}\label{eq:distancemetric}
d_{\theta}((L_1,\alpha_1),(L_2,\alpha_2)) = \sqrt{v(L_1,L_2)^2+ (\theta \frac{d_{\circ}(\alpha_1,\alpha_2)}{180^\circ})^2},
\end{equation}

We denote the heading penalty parameter through $\theta$.  Intuitively, we want to penalize points that are very close to each other based on location but has diametrically opposite direction. This is often the case for parallel roads where each lane corresponds to the traffic in opposite directions.

\begin{lemma}
The distance metric $d_{\theta}(.,.)$ is a metric.
\end{lemma}
Proof. This follows directly from Cauchy-Schwarz inequality and the fact that both $v$ and $d_{\circ}$ are metrics:
$$(d_{\theta}((L_1,\alpha_1),(L_2,\alpha_2))+d_{\theta}((L_2,\alpha_2),(L_3,\alpha_3)))^2=$$
$$=v(L_1,L_2)^2+ (\theta \frac{d_{\circ}(\alpha_1,\alpha_2)}{180^\circ})^2 + v(L_2,L_3)^2+ (\theta \frac{d_{\circ}(\alpha_2,\alpha_3)}{180^\circ})^2 + $$
$$2\sqrt{v(L_1,L_2)^2+ (\theta \frac{d_{\circ}(\alpha_1,\alpha_2)}{180^\circ})^2}\sqrt{v(L_2,L_3)^2+ (\theta \frac{d_{\circ}(\alpha_2,\alpha_3)}{180^\circ})^2}\geq$$
$$v(L_1,L_2)^2+ (\theta \frac{d_{\circ}(\alpha_1,\alpha_2)}{180^\circ})^2 + v(L_2,L_3)^2+ (\theta \frac{d_{\circ}(\alpha_2,\alpha_3)}{180^\circ})^2 + $$
$$2v(L_1,L_2)v(L_2,L_3)+ 2\theta \frac{d_{\circ}(\alpha_1,\alpha_2)}{180^\circ}\cdot \theta \frac{d_{\circ}(\alpha_2,\alpha_3)}{180^\circ}=$$
$$(v(L_1,L_2)+ v(L_2,L_3))^2+(\theta \frac{d_{\circ}(\alpha_1,\alpha_2)}{180^\circ}+\theta \frac{d_{\circ}(\alpha_2,\alpha_3)}{180^\circ})^2 \geq$$
$$v(L_1,L_3)^2+ (\theta \frac{d_{\circ}(\alpha_1,\alpha_3)}{180^\circ})^2=d_{\theta}((L_1,\alpha_1),(L_3,\alpha_3))^2$$.

\subsection{Densification}
\label{subsec:densification}
An important pre-processing step, before inferring the nodes of the graph, is data densification that addresses two challenges: (a) different sampling rates and (b) data disparity. Due to power concerns, GPS measurements might not be reported at an adequate and constant rate. Similarly, based on the speed limits, two GPS points can be far away from each other. At a sampling rate of one GPS point every 10 seconds, for a car driving at 60 $kmph$, two consecutive points could be as far as 170 meters even without GPS noise. This issue is especially pronounced when there is data \emph{disparity}. Specifically, in low-volume road segments (e.g., residential areas), the amount of data available is limited, leading to potentially missing those road segments. 

The intuition behind densification is to introduce artificial GPS points along the trajectory of the moving vehicle in the following manner. 
If the two consecutive GPS points $x_1 = (lon_1,lat_1,\alpha_1)$, $x_2= (lon_2,lat_2,\alpha_2)$ on the trajectory have headings with angular difference of 
less than a threshold (say $5^{\circ}$), we add $sr$ GPS points points between them in an equidistant manner.
$sr$ can be application and dataset dependent.
For our paper, we set it as $s = \lfloor v(L_1,L_2)/ 20m \rfloor$. 
By doing this we effectively create a point every (approximately) $20m$ along trajectories on the straight roads. 
We would like to note that densification is an optional step that can be skipped if the data is dense enough, 
or the map accuracy in low-frequency roads is not important for the end-user of the map. 
Since, our design objective is to achieve high levels of map accuracy for both major and minor roads,
we densify the original data before clustering. 

%Overall our densification process roughly doubles the number of datapoints.

%In addition to densification we identified a single vehicle in our dataset which produced abnormal location infromation (lat/lon shifted northeast by several hundred meters) and filtered it out of the dataset. 

\subsection{Seed Selection and Cluster Assignment}
\noindent {\bf Seed Selection:}
After the optional densification step, we cluster the GPS data points in ($T$) while ignoring the trajectory information. 
The cluster centers obtained will form the nodes $V$ of the inferred road network $G$ and are then connected using the trajectory information.
In this paper, we use $k$-Means algorithm for clustering that has been shown to work well in prior work even without usage of the angle information~\cite{agamennoni2011robust,schroedl2004mining}.
We use the following process for selecting a good set of seed nodes.
%our input is transformed to the list of records in the form: $(user_-id, timestamp, lon, lat, heading)$. Our goal in this secion is to cluster together similar data points which fall in the same road segment. For clustering purposes we look at each data point independently; thus we do not use the trajectory information (inferred from the timestamps and $user_id$s) to cluster the points. These clusters will later be tied together using the trajectory information in the next section to form the road network. 

%Our clustering is $k$-means with several modifications which we will detail below. The distance metric takes into account the location $L=(lat,lon)$ of the datapoint and the heading. For two points $(L_1,\alpha_1)$ and  $(L_2,\alpha_2)$ the distance is
%\begin{equation}\label{eq:distancemetric}
%d((L_1,\alpha_1),(L_2,\alpha_2)) = \sqrt{v(L_1,L_2)^2+ (\theta d_{\circ}(\alpha_1,\alpha_2))^2},
%\end{equation}
%where $v(L_1,L_2)$ is Vincenty distance in meters,  $d_{\circ}$ is the angle distance
%\footnote{$d_{\circ}(\alpha_1,\alpha_2) = min(|\alpha_1-\alpha_2|,360^\circ-|\alpha_1-\alpha_2|)$.}
%and $\theta$ is the heading penalty parameter. For most of the analysis in Sec \ref{} we set $\theta$ in such a way to penalize $180^\circ$ heading difference equivalently to $50m$ of geo distance. 

%It is straighforward to check that the above distance measure satisfies the triangle inequality and is indeed a metric and its units are meters. \textbf{RS:Note to myself: we should evaluate different $\theta$ too!}

We start with an empty seed list and go through the GPS points sequentially. 
A GPS point $(L,\alpha)$ is added to the seed list if there are no other seeds within a specified radius $seed_-radius$ using the distance function $d_{\theta}(\cdot)$.
This process ensures that every point is within a distance $seed_-radius$ to some seed point.
Additionally, this also results in the seeds being uniformly spread throughout the space of the map by making sure there are no two seeds within a distance $seed_-radius$ of each other. The parameter $seed_-radius$ determines the density of the clusters. For example using a small $seed_-radius$, say $3m$ can result in inferring each lane in a highway as an independent road segment, while choosing large $seed_-radius$ may result into merging points from close (parallel) streets into the same cluster. 
%In Section \ref{sec:evaluation} we examine the impact of $seed_-radius$ on the overall quality of the maps. 
%\textbf{RS:Note to myself: we should evaluate different $seed_-radius$!}

\noindent {\bf $k$-Means Clustering:} 
After the initial cluster centers are selected, we run the standard $k$-Means algorithm (where $k$ is the number of seeds selected).
In the assignment step, a GPS point is assigned to the closest centroid using the distance metric $d_{\theta}$ defined in Section~\ref{subsec:distanceMetric}.
In the update step, the cluster centroid is updated using standard mean along the $lat/lon$ coordinates and the mean of circular quantities for the heading coordinate. Formally, the centroid for a cluster of points $S_{i}$ is:
$$m_{i} = (\overline{lat},\overline{lon},\overline{\alpha})$$ 
where:
$$\overline{lat} = \frac{1}{|S_{i}|}\sum_{x \in S_{i}} x.lat.\ \ \ \  
 \overline{lon} = \frac{1}{|S_{i}|}\sum_{x \in S_{i}} x.lon $$
and 
$$\overline{\alpha} = atan2(\frac{1}{|S_{i}|}\sum_{x \in S_{i}} \sin x.\alpha ,\frac{1}{|S_{i}|}\sum_{x \in S_{i}} \cos x.\alpha).$$

It is well known that $\overline{\alpha}$ is a maximum likelihood estimator of the Von-Mises distribution (the spherical Gaussian) \cite{fisher1995statistical}.

\begin{comment}
\textbf{RS:As I was writing i realized that mean od circular quantities may or may not (more likely not) be the least-square estimator for the angle distance metric $d_{\circ}$. Something to check!}. We iterate $k$-means until the cost reduction becomes smaller than 0.0001 times cost. The final result of the $k$-means is a set of $k$ clusters which represent different road segments. Some clusters represent straight road segments and among them the variability of the heading angles among the points within the cluster is indeed very small (around $1-2^\circ$). However, clusters around intersections and roundabouts contain points with a wider range of angles which may confuse the road creation process (Section \ref{}). Hence we have the final step in the clustering phase:
\end{comment}

\noindent \textbf{Cluster Splitting}: 
After the clustering is done, we perform a post-processing routine to ensure that GPS points in each cluster are homogeneous.
Specifically, if we observe that the heading angles in a cluster have a large variance, we proceed by splitting the cluster 
into smaller ones that are individually more homogeneous.
This additional step is necessary to handle some complex intersections such as roundabouts that present a wide range of angles.
Splitting clusters ensures that our map can naturally handle complex intersection shapes.

For a cluster $v_i$ with mean heading $\overline{\alpha_i}$ we define heading variability $hv_i$ as the average of $d_{\circ}(x.\alpha,\overline{\alpha_i})$, across all points $x$ within the cluster\footnote{Note that angular arithmetics do not allow us to use standard measures of variability such as standard variation or variance.}.
If the heading variability $hv_i$ is greater than some threshold (say, $10^\circ$) 
we split $v_{i}$ into two clusters by partitioning $v_i$ based on the angle information of its points.
We then recalculate the centroids of the new clusters accordingly.

\subsection{Edge Inference} \label{sec:graphConstruction} 
In the previous subsection we used the clustering process to infer the nodes $V$ of the output graph. 
In this subsection,  we seek to infer the candidate set of edges $E_C$ of $G$ by using the trajectory information. 
We will then prune $E_C$ by constructing a spanner of the graph $G=(V,E)$.

\begin{comment}
The clustering described in the previous section took the GPS point cloud and identified the ``skeleton'' of the eventual road network.
In this section, we describe how to utilize the trajectory information to convert the skeleton into a routable directed graph. 
Note that it is not possible to construct the map only from the skeleton.
While simple structures such as straight road segments could be easily identified, complex intersections such as roundabouts are hard to infer without trajectory information.
One of the major contributions of the paper is a principled approach to construct a sparse graph that succinctly summarizes the trajectories.
The map construction proceeds in two stages.
We begin by constructing a cluster connectivity graph that represents the trajectories and then sparsify it to obtain a more compact graph.
\end{comment}

The output of the clustering stage is a set of cluster centroids $\mathcal{V} = \{v_1, v_2, \ldots, v_k\}$
where each centroid is a triple of $\langle lat, lon, a\rangle$. 
We now construct a cluster connectivity graph $G_C = (V, E_C)$ that integrates the clustering information and trajectory information.
%where each centroid is a triple of $\langle latitude, longitude, angle\rangle$.  
%\textcolor{red}{ There are two types of edges - nearest neighbor based and trajectory based.}
%A (nearest neighbor based) directed edge $e_{i,j}$ is added between nodes $v_i$ and $v_j$ (corresponding to clusters $c_i, c_j$) if 
%$c_i$ is the nearest cluster to $c_j$ and they are in the same angle.
To create the edges we use the set of densified trajectories $T_D$  as follows:
%In the next step, we process the set of trajectories one at a time and convert the information into a series of trajectory based edges.
Consider a trajectory $tr=\{x_1, x_2, \ldots, x_{|tr|}\}$.
We transform the $tr$ from a sequence of GPS points into a sequence of cluster centroids $tr_v = \{v_{i,1}, v_{i,2}, \ldots, v_{i,|tr|}\}$
where $v_{i,j}$ is the closest centroid to point $x_i \in tr$.
For each pair of consecutive centroids $ (v_{i,j}, v_{i, j+1}) \in tr_c$, we add an edge between them if the following conditions are satisfied:

\noindent \textbf{Self Loop Avoidance:} $v_{i,j} \neq v_{i, j+1}$. If two consecutive data points fall into the same cluster, then we do not add a self edge as it does not provide any additional information.

\noindent \textbf{Spurious Edge Avoidance:} 
\[\scriptstyle f_e(v_{i,j},v_{i, j+1})\geq \max( 1, \ln( \min( f(v_{i,j}), f(v_{i, j+1}))) -1)
\] 
where $f(u)$ is the number of datapoints assigned to cluster $u$ and $f_e(u,v)$ number of trajectories which draw the edge $(u,v)$ in the cluster connectivity graph $G_C$. This condition eliminates spurious edges which may be created due to noise or abnormal vehicle behaviour (e.g. going in the wrong direction in a one-way street). 

\begin{comment}
\item \textbf{Collinearity:} 
\[\scriptstyle \max(d_\circ(\overrightarrow{c_{i,j}c_{i,j+1}},\alpha_1)),d_\circ(\overrightarrow{c_{i,j}c_{i,j+1}},\alpha_1)))< 4*(d_\circ(\alpha_1,\alpha_2)+min(hv_1,hv_2))\]
 This condition is enforced almost exclusively on clusters representing road segments on two parallel streets going side-by-side and seeks to eliminate edges between clusters corresponding to two different lanes. The  $\overrightarrow{c_{i,j},c_{i,j+1}}$ is direction of the line segment between cluster
    centroids $c_{i,j}$ and $c_{i,j+1}$, $\alpha_j$ is the heading of the clusters $c_j$ and the $hv_j$  is the heading variability of cluster $c_j$ defined as mean angular difference between $\alpha_j$ and the heading among all data points assigned to cluster $c_j$ ($j=1,2$).
\end{comment}

\subsection{Graph Sparsification through Spanners}

\begin{algorithm}[h]
\caption{Two-Phase \kha Algorithm}
\label{alg:twophase}
\begin{algorithmic}[1]
\STATE{\bf Input:} A collection $T$ of trajectories, $\alpha$
\STATE{\bf Output:} A directed planar graph $G=(V,E)$.
\STATE {\bf Phase 1:}
	\STATE \qquad {\em (Optional)} $T_D \leftarrow$ Densification$(T)$ 
	\STATE \qquad $G_D \leftarrow$ DistinctPoints($T_D$)
	\STATE \qquad Let $S \leftarrow$ SelectSeeds($G_D$)
	\STATE \qquad $V \leftarrow$ $k$-Means($|S|$, $G_D$)
\STATE  {\bf Phase 2:}
	\STATE \qquad Let $E_C \leftarrow$ EdgeAssignment$(V,T_D)$
	\STATE \qquad $G(V, E^\prime) \leftarrow$ Spanner$(V,E_C,\alpha)$
	\STATE \qquad $G(V, E) \leftarrow$ Duplexification$(G(V, E^\prime))$
\end{algorithmic}
\end{algorithm}

The inferred graph $G=(V,E_C)$ from the previous section has potentially several redundant
edges which makes $G$ unusable for routing. 
%The cluster connectivity graph $G_C$ encoded the trajectory information and the nearest neighbor information for each cluster.
%Let us now consider the deficiencies of $G_C$.
%First, it contains a cacophony of edges - in the worst case, one edge per each consecutive GPS points for each  trajectory in the collection.
%This makes $G_C$ totally unusable for routing purposes.
%\textcolor{red}{Figure~\ref{TBD} shows an example of cluster connectivity graph for a popular roundabout in Doha, Qatar}.
Recall from Section~\ref{subsec:densification} that for a car driving at 60 $kmph$ (around 40 $mph$), 
two consecutive points obtained ten seconds apart could be as far as 170 meters (560 feet.) 
However, due to the densification step, the cluster centroids could be as near as 20m (65 feet).
So these two points could create a very long edge which may not follow the underlying road geometry.

Our objective is to construct a sparsified graph that retains the same connectivity information but also connects clusters that are near each other maintaining the road shapes.
Consider a trajectory based edge $(v_i, v_j) \in E_C$. This implies that $v_j$ is reachable from $v_i$ through the underlying road network.
However, due to the sampling rate, they might not necessarily be adjacent to each other.
Hence, we would like to identify a smooth path between $v_i$ and $v_j$ that connects them through clusters that are near each other. 
Let us now formalize this problem.
Given candidate graph $G=(V, E_C)$, our objective is to obtain a sparsified graph $H = (V, E)$ where:

\noindent  $E \subseteq E_C$ . i.e. $H$ is a subgraph of $G$

\noindent $\forall e = (v_i, v_j) \in E_C$, $d_{H}(v_i, v_j) \leq \alpha \times d_{G}(v_i, v_j)$. This constraint ensures that the sparsification preserves approximate distance between each pair of vertices in $G$. In other words, for any pair of vertices $(v_i, v_j) \in V$, their distance $d_{H}(v_i, v_j)$ in $H$ is at most $\alpha$ times their distance $d_{G}(v_i, v_j)$ in $G$. Setting larger values of $\alpha$ generates sparser graphs. Throughout the paper we use $\alpha = \sqrt{2}\approx 1.41$ which removes the cross edges in the orthogonal streets.

This problem can be abstracted as a well studied combinatorial problem of graph spanners \cite{narasimhan2007geometric}.
Given a graph $G=(V,E_C)$, a subgraph $H=(V,E)$ is called a $\alpha$-spanner of $G$, if for every $u,v \in V$, 
the distance from $u$ to $v$ in $G$ is at most $\alpha$ times longer than the corresponding distance in $G$.
There has been extensive prior work on developing efficient algorithms for spanner construction.
For the purpose of our paper, we use a simple greedy spanner algorithm depicted in Algorithm~\ref{alg:greedySpanner}.
A naive implementation has a time complexity of $O(n^3 \log n)$ while it can be improved to $O(n^2 \log^2 n)$ 
using advanced data structures \cite{narasimhan2007geometric}.
However, we empirically observe that by not recomputing the paths we can obtain a sub-quadratic runtime complexity while retaining full connectivity.

The final step in \kha is 'duplexification' of slow road segments. Namely, certain low-tier streets receive very few data points, and often in only one direction which may confuse the \kha to consider it as a one-way street. We make sure that these capillary roads are two way by adding an edge $(v,u)$ to the final graph $H$ if: $(u,v)\in H$ and maximum observed speed of all data points assigned to $u$ and $v$ is $\leq 60kmph$.

\begin{algorithm}
    \caption{Greedy Spanner Algorithm}
    \label{alg:greedySpanner}
    \begin{algorithmic}[1]
		\STATE {\bf Input:} $G = (V, E_C)$, $\alpha$
        \STATE $E \leftarrow \emptyset$ , $H = (V, E)$
        \FOR {each edge $(v_i, v_j) \in E_C$ in the order of decreasing weight}
            \IF {$d_G(v_i, v_j) > \alpha d_H(v_i, v_j)$}
				\STATE $E \leftarrow E \cup (v_i, v_j)$
			\ENDIF
        \ENDFOR
        \RETURN $H$
    \end{algorithmic}
\end{algorithm}

%\textcolor{red}{In Figure \ref{TBD} we show the results of both $G_C$ and $G$ for a popular intersection called TV roundabout in Doha, Qatar}.

The pseudocode for \kha is depicted in Algorithm~\ref{alg:twophase}.

\section{\khastar: Online Algorithm}
\label{sec:onlineAlgorithm}

In this section, we present \khastar, an online algorithm that can update the map as new trajectories arrive.
In this algorithm, the two phases of the offline algorithm - clustering and sparsification - are combined into one phase.
This enables us to process trajectories that arrive in streaming fashion.

{\bf Streaming Setting:}
We consider the following streaming setting: 
our algorithm is provided a pair of GPS data points, $x_i, x_{i+1}$, that are taken consecutively.  

Our algorithm is generic enough to handle various GPS streaming models that have been previously studied.
As an example, it generalizes trajectory streaming model where the algorithm is provided trajectory at a time.
Given a trajectory $tr$, one can convert it to our streaming model by considering every consecutive pair of points.
Similarly, our model is equivalent to the sliding window streaming model with the widow size of 2. 
Our model cannot directly handle the GPS navigational stream format where points arrive one at a time.
However, with a simple modification, where we store the previous data point from every unique source, we can still apply our online algorithm.

\noindent {\bf Challenges:}
Recall that our offline algorithm has two phases: (i) clustering and (ii) graph sparsification.
Both of them required access to the entire data.
The clustering phase consists of seed selection and actual $k$-Means.
The graph sparsification required to order the edges of the graph in the decreasing order of weight.
However, in the streaming setting, we have to process each pair of GPS points immediately and it is not feasible to conduct expensive operations. 
Our solution combines online adaptation of $k$-Means clustering algorithm and online spanner algorithm.

\noindent {\bf Online Algorithm for $k$-Means and Spanners:}
We begin by providing a brief intuition behind prior work on online $k$-Means algorithm \cite{DBLP:conf/alenex/2016} that was published last year.
In this problem, the data points arrive one at a time and the objective is to provide a clustering 
that is competitive with the offline variant of $k$-Means that has access to all the data points.
The algorithm assigns the first $k+1$ as cluster centroids.
When a new data point arrives, it is assigned to the nearest centroid if the distance is less than some threshold $f_i$.
If not, the data point becomes a new cluster on its own.
As the number of clusters becomes larger, the threshold is periodically doubled thereby reducing the likelihood that a new cluster is created for new points.
The online version of spanner algorithm is adapted from \cite{McGregor:2014}.
Given an $\alpha$-spanner graph $G$ and a new edge $e=(u,v)$, we add the edge to $G$ if $\alpha \times w(e)$ is less than the distance $d_G(u,v)$.

\begin{figure}[h]
\centering
\includegraphics[width=\linewidth]{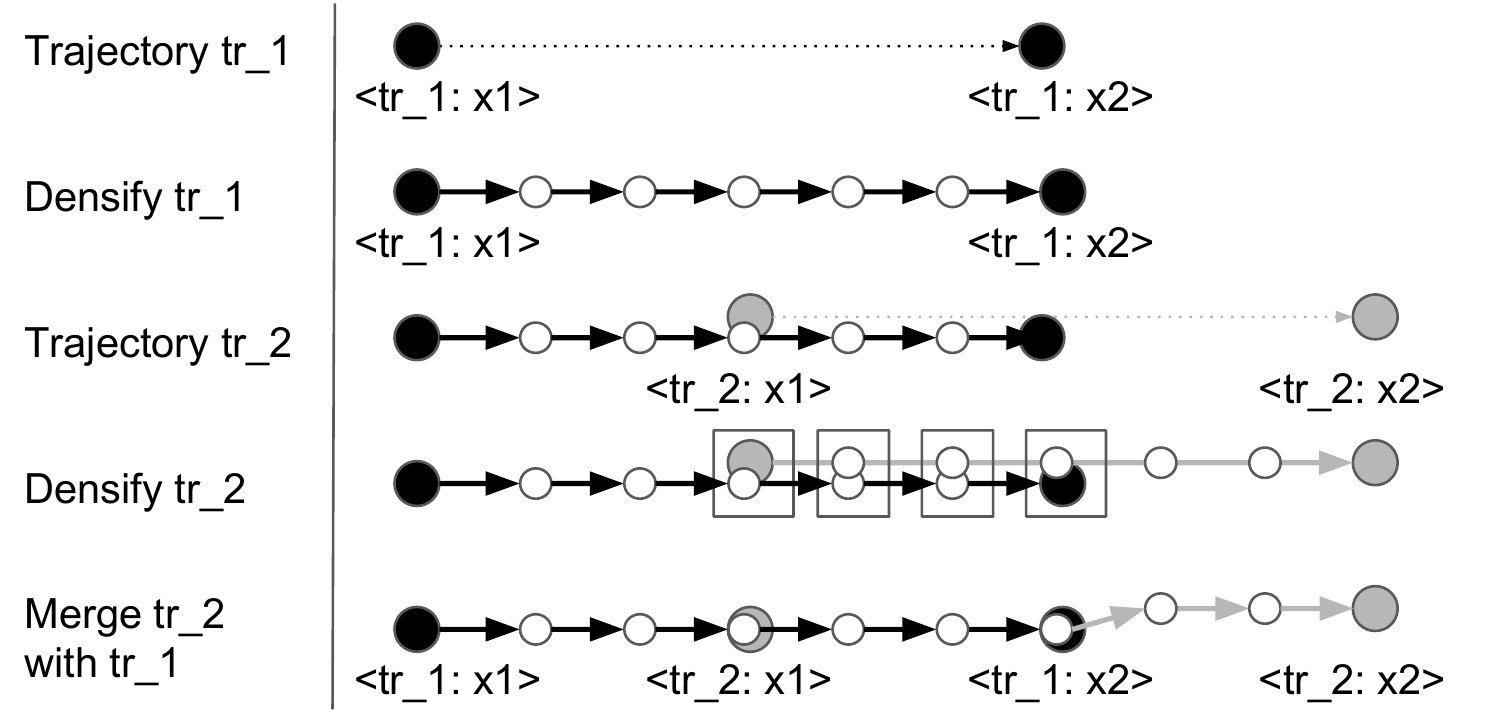}
\caption{Example of streaming densification and merging of trajectories. The first trajectory $tr_1$ is densified and all its points are considered as new clusters, whereas the densification of the second trajectory $tr_2$ results in the assignment of some of its points to the clusters generated for $tr_1$}
\label{fig:onlineAlgoSteps}
\end{figure}

\noindent {\bf Online Map Construction Algorithm -- \khastar:}
We begin with an empty graph $G=(V,E)$.
When a pair of GPS points $(x_i, x_{i+1})$ arrives, we first densify them by creating a set of equidistant points 
$P=\{p_1, \ldots, p_l\}$ where $p_1=x_i$ and $p_l=x_{i+1}$ and each consecutive pair of points differ by a fixed densification threshold of $sr$ meters.
Each point $p_i$ is assigned to the closest node (cluster) $v^{*} \in V$ if it is within a radius distance $cr$ and 
the difference in angle is less than some heading angle tolerance $ha$.
If not, we create a new node and assign $p_i$ to it.
Next, the algorithm checks whether an edge should be created from node $v^{*}_{prev}$ (which correspond to the node to which the point $pt_{i-1}$ was assigned) and nodes $v^{*}$. This edge is created if and only if the angle difference  between $v^{*}_{prev}$ and $v*$ as well as that between $v^{*}_{prev}$ and the vector $\overrightarrow{(pt_i, pt_{i+1})}$ are both lower or equal to $ha$.
In addition, we also ensure that the graph remains an $\alpha$-spanner.
Given the online and streaming nature of \khastar, we had to give up on the angular distance $d_{\theta}(.,.)$ and replace it with a hard angle based threshold filtering to better control for angle differences. 
Figure~\ref{fig:onlineAlgoSteps} shows how the online algorithm proceeds step by step.
Algorithm \ref{alg:onlinemapinference} provides the pseudocode for \khastar.

\begin{algorithm}
    \caption{\khastar for online map inference}
    \label{alg:onlinemapinference}
    \begin{algorithmic}[1]
	\STATE {\bf Input:} Trajectory $tr = \{x_1,x_2, \ldots x_{|tr|}\}$, Road graph $G = (V, E)$, that can be empty 
	\STATE {\bf Parameters:} clustering radius ($cr$, in meters), sampling rate ($sr$, in meters), 
		heading angle tolerance ($ha$, in degrees)
        
        \FOR {each consecutive pair of points $e(x_i, x_{i+1}) \in tr$}
	  \STATE $tr_D$ $\leftarrow$ Densification(($x_i, x_{i+1}$), $sr$)
	  \FOR {each point $p_i \in tr_{D}$}
	    %\STATE $v^* = argmin(distance(p_i, v_i) | v_i \in V ) | distance(p_i, v^*) \leq c_r \vee angle\_diff(p_i, v^*) \leq h_a$
        \STATE $v^*$ = $argmin(d(p_i, v_i) | v_i \in V )$
        \IF {$v^{*}$ is within the distance tolerance $cr$ and angle tolerance $ha$}
            \STATE Assign $p_i$ to cluster $v^{*}$
	    \ELSE
            \STATE Create a new node $v^{*}$ for $p_t$
            \STATE $V \leftarrow V \cup \{v^{*}\}$
	    \ENDIF
	\IF {$d_{G}(v^*_{prev}, v^*) > \alpha \times d(edge(v^*_{prev}, v^*))$}
	    \STATE $E \leftarrow E \cup \{(v^*_{prev}, v^*)\}$
	\ENDIF
	    \STATE $v^*_{prev} \leftarrow v^*$ 
	    \ENDFOR
        \ENDFOR
        \RETURN $G$
    \end{algorithmic}
\end{algorithm}

\noindent {\bf Incremental Map Maintenance:}
A key advantage of the online algorithm \khastar is that it can be easily modified so that both the geometry and connectivity is updated based on the traffic patterns.
Let us consider the major scenarios.

\noindent 1) {\em New Road:} No special processing is needed. As we observe new trajectories from new roads, the map will be automatically updated. 

\noindent 2) {\em Shifting of an Existing Road:} There are two cases. If there is only a slight shift (of less than $sr$), then our algorithm will not detect it. 
            However, $sr$ is often set to a low value and hence the shift might not have a major impact in routing.
            If the shift is major, we will create a parallel road based on the traffic observed on the new road. 
            A time-stamp indicating the last time a cluster was visited is used to decide whether the cluster should be disabled. We can use a time decaying function to periodically eliminate those nodes and edges that have not seen any new traffic recently.
            Of course, this threshold has to be based on the prior traffic over the nodes and edges.
            In other words, major roads must be marked inactive while we provide some leeway for roads with low traffic.

\noindent 3) {\em Road closure and blocking:} 
            Our method can handle both temporary and permanent closure of the roads using the same idea as above.
            We store with each node and edge the time-stamp of the last GPS point that got assigned to it and the ``typical'' travel pattern.
            If the delay is large, we mark the edge as ``inactive''.
            If the closure is temporary, a road is marked as active.

\noindent 4) {\em Shortest Distance Maintenance:}  
            Due to the online spanner algorithm, there might eventually be multiple paths between two nodes $(u,v)$.
            We periodically invoke the offline spanner algorithm to ensure that the graph is as sparse as possible and remove the non-shortest paths between them.

\section{Evaluation}
\label{sec:evaluation}

We discuss in this section the results of comparing our proposed solution \kha to state of the art map construction algorithms. The baseline algorithms are selected in a way to cover all major approaches for map inference, namely Edelkamp~\cite{edelkamp2003route} for $k$-means based techniques, Cao~\cite{cao2009gps} for trace merging based techniques, Biagioni~\cite{biagioni2012inferring} for KDE based techniques, and Chen~\cite{ChenLHYGG16} which is the closest to our work in that it uses data with similar features to ours (e.g., angle, speed.) In the following, we first introduce the datasets and evaluation metrics used. Next, we report the comparison of \kha to state of the art algorithms. Finally, we assess the robustness of our solution to different parameter initialization settings.

\subsection{Data and methodology}
\begin{table}[ht]
\small
\caption{Characteristics of Datasets}
\begin{center}
\begin{tabular}{|c|c|c|c|c|}
\hline
\textbf{Dataset} & \# Days & \# GPS points & \# Vehicles  & Covered roads (km)\\ 
\hline
Doha & 30 & 5.5M & 432 & 300\\
\hline
UIC & 29 & 200K & 13 & 60 \\
\hline
\end{tabular}
\end{center}
\label{tab:DatasetChar}
\end{table}

As we discussed earlier, our map inference process uses data generated by a fleet of vehicles with GPS-enabled devices. In this paper we utilize two datasets from Doha (Qatar) and UIC (Chicago) with basic statistics reported in Table \ref{tab:DatasetChar}. 

The two datasets are different in that \textit{Doha} dataset reports  the \textit{speed} and the \textit{heading} of the moving direction in addition to the  \textit{location} $(lon,lat)$ of the vehicle, while \textit{UIC} datasets reports only the \textit{location}. Since \kha utilizes both speed and heading information we infer them from the consecutive data points. Heading is measured in angles against the North axis in degrees reporting values from $0$  to $360^{\circ}$. 

\textit{Doha} dataset covers a rectangle (in $lat,lon$ coordinates) of 6km$\times$8km in an urban region in the city of Doha with a mixture of highways, high and medium volume roads, capillary streets, and roundabouts. The \textit{UIC} dataset covers an area of approximately 2km$\times$3km in downtown Chicago and is generated by a fleet of University buses with relatively regular routes. 

We preprocessed both datasets to eliminate those datapoints with speed $\leq 5kmph$ which are known to have non-trivial noise when reporting location. 

\kha has two main parameters. In Doha dataset we set $seed_-radius = 20m$ to cover 3-lane roads with minimal noise, while in UIC dataset we set $seed_-radius = 80m$ as we observe much higher noise levels, which go as high as $60m$ for a single road segment. In both cases we use $\theta=2\cdot seed_-radius$ to ensure that two datapoints from the same street going into oposite directions are not assigned to the same cluster. We also experimented with other choice of parameters and observe low sensitivity of final maps to the parameter choice. However, a more detailed examination of parameter space is left for future work. 

Evaluating the quality of an automatically generated map is a challenging question. The de-facto standard for measuring the quality of the map is the holes-and-marbles method introduced by Biagioni and Eriksson \cite{biagioni2012inferring} which we describe now. 

\textbf{Holes-and-marbles methods.} Biagioni and Eriksson \cite{biagioni2012inferring} proposed the following two methods for measuring the accuracy of the inferred map. 

GEO method evaluates how well the given map geometrically matches a ground truth map. Throughout the paper we use the OpenStreetMap (OSM) snapshot of the same region as the ground truth map\footnote{As far as we can tell several intersections are not accurately represented by the OSM in our region of interest due to construction works, but this has relatively small impact on the matching scores.}. GEO method samples points every 5 meters from both maps and puts a marble in each sampled point of the tested map and puts a hole in each sampled point of the ground truth map. We say that marble (hole) is matched if there is a hole (marble) within $matching_-threshold$. We vary this threshold in the range of $5m-30m$ and evaluate $precision$, $recall$ and $f_{score}$ as follows:
$$precision = \frac{\# matched_-marbles}{\#all_-marbles},\ \ recall = \frac{\# matched_-holes}{\#all_-holes}$$
$$f_{score} = 2\frac{precision\cdot recall}{precision + recall}$$

TOPO method evaluates the topological characteristics of the map. As it was done in other map-evaluation studies we eliminate the edges of the ground-truth map without a single trajectory passing through it. This way TOPO method measures the topological quality of the map: how accurately the inferred map assesses the connectivity structure of the inferred roads. Namely from a randomly sampled marble (hole) we first find all the marbles (holes) which could be reached from the starting point within a $radius$; throughout the paper we set the $radius = 2000m$. For the neighborhoods of each sampled starting point we calculate $f_{score}$ as above and report the mean $f_{score}$ over a sample of 200 randomly selected starting points. Note, that starting marble and starting hole belong to two different maps and virtually never match exactly; hence we enforce them to be within $1m$ distance and belong to the roads with the same direction (within a small angle difference). 

\subsection{Comparison with state-of-the-art}

In this section we compare \kha with state of the art (SOTA) map-inference algorithms: Cao \cite{cao2009gps}, Edelkamp \cite{edelkamp2003route}, Biagioni\cite{biagioni2012inferring}, and Chen \cite{ChenLHYGG16}. For Cao, Biagioni and Edelkamp we use the source code developed and provided by the authors, while for Chen we use our implementation of the algorithm as the original implementation was not made available to us. 

An important characteristics of some of the existing solutions is poor scalability. Namely it takes 3,558 and 6,621 seconds to run Cao/Edelkamp on a single day worth of data ($\approx 200K$ data points), with only one iteration of Cao's clarification step. Hence we report the comparison of the different algorithms using one day worth of data from Doha. Furthermore, trying to scale these algorithms to one week of data has resulted in ``out of memory'' problems in the machines we used for the evaluation.  

\subsubsection{GEO Comparison}
\begin{figure}%[ht]
  \includegraphics[width=0.33\textwidth]{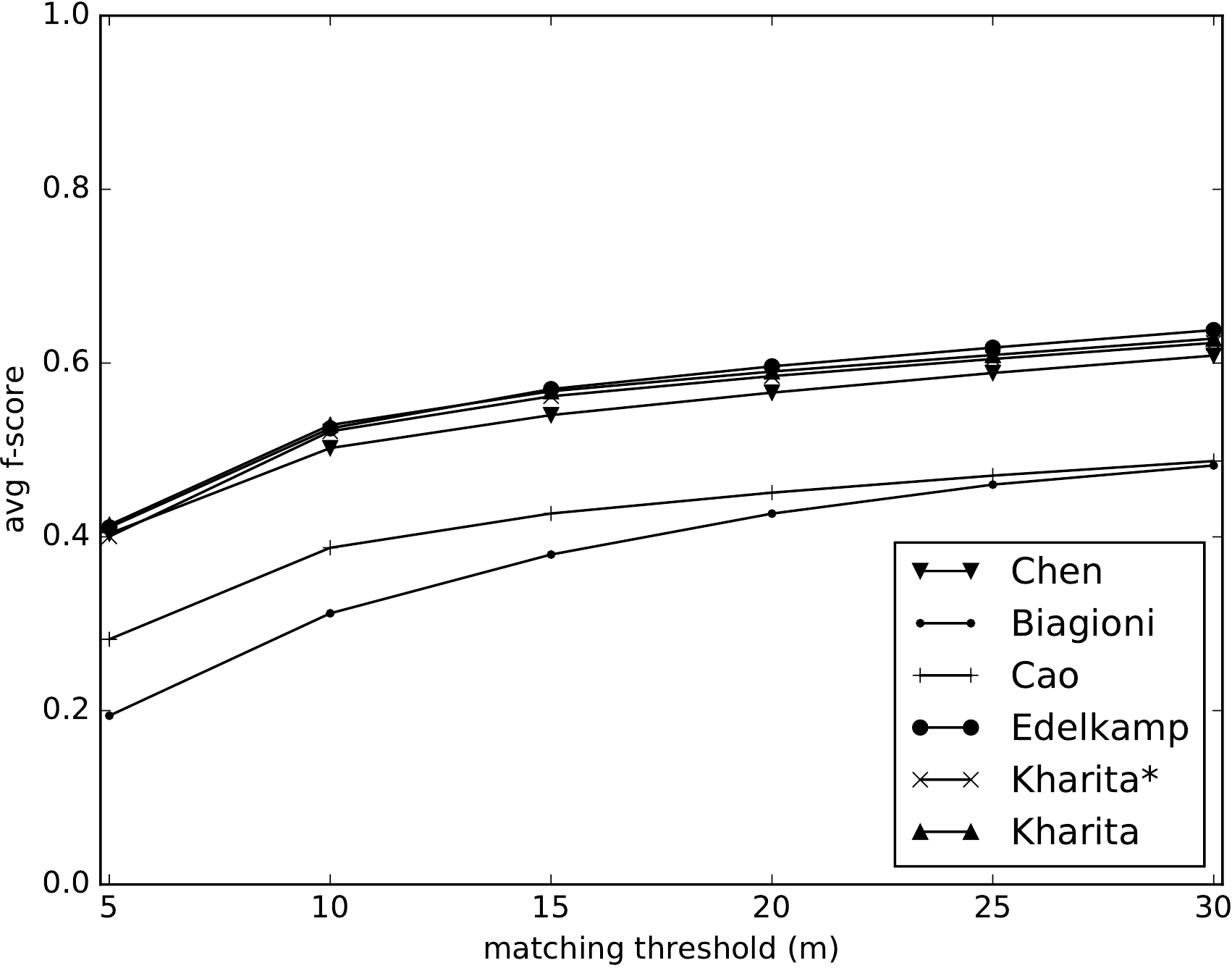}
  \includegraphics[width=0.33\textwidth]{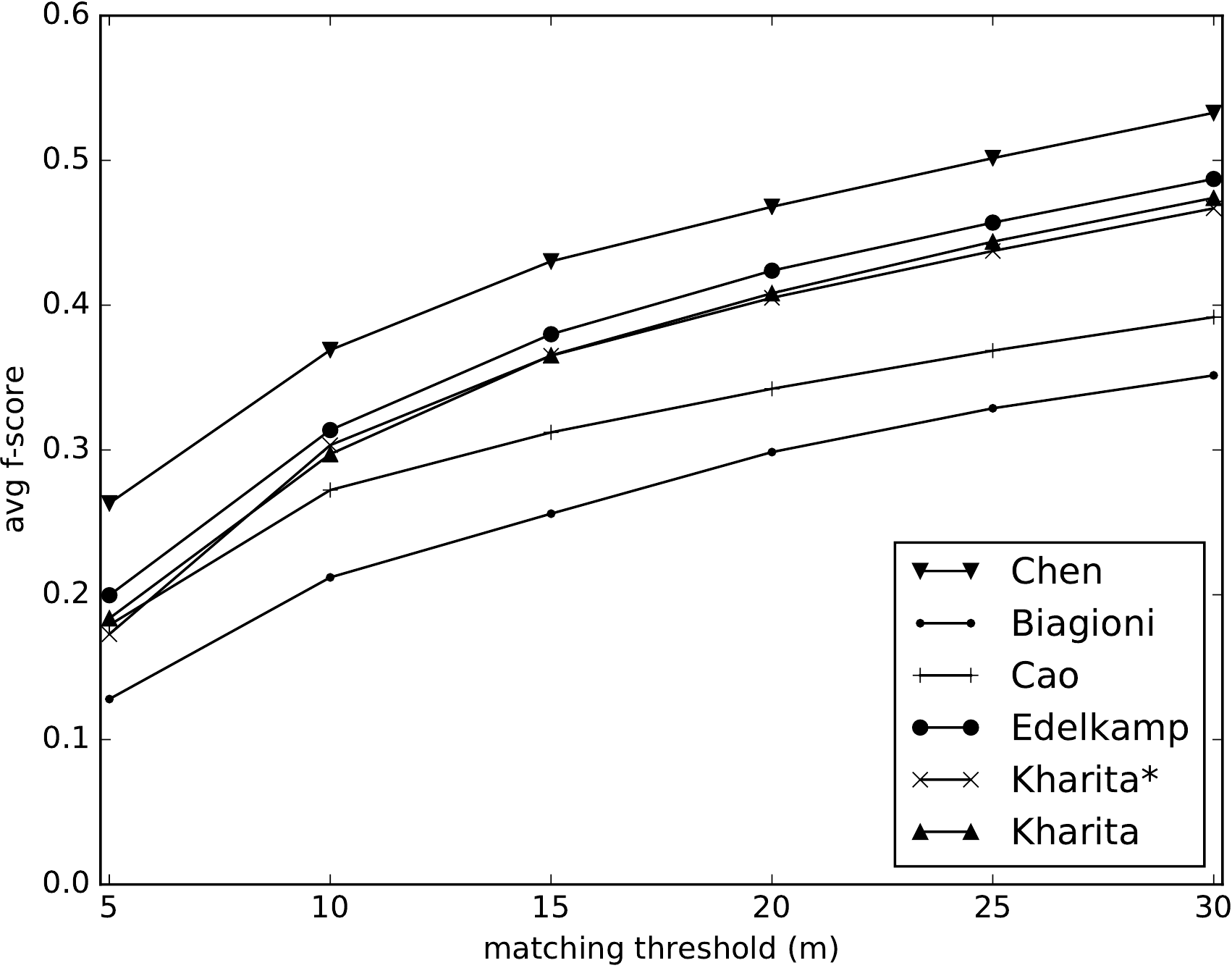}
  \caption{GEO $f_{score}$ for different map-inference solutions. (top) Doha (1-day), (bottom) UIC data}
  \label{fig:GEO_SOTA}      
\end{figure}

In Figure \ref{fig:GEO_SOTA} we report the GEO $f_{score}$ for the four map inference solutions, measured against the underlying OSM map in the rectangles of interest for both Doha and UIC datasets. 
%They measure how well is the underlaying map covered by the inferred map. 
%Note that $f_{score}$ of, say, 0.5 suggests that around 50\% of the road network is correctly inferred by the algorithm. 
The GEO $f_{score}$ is predominately determined by the amount of the data and their coverage of the underlying road network, but also depends on the edge-creation process. Namely some algorithms are more conservative in the way they infer a road segment and hence have lower $f_{score}$; Cao and Biagioni belong to this category. Other algorithms require a small number of trajectories to infer a particular road segment and consequently have larger $f_{score}$. Note that difference in $f_{score}$ among those (Chen, Edelkamp, \kha, \khastar) is relatively small which is a result of the fact that geometrically each one of them closely covers the street segments which have at least one trajectory passing by. However, GEO method is oblivious to the connectivity structure of the road network which is captured by TOPO method. 

\subsubsection{TOPO Comparison}
\begin{figure}
   \includegraphics[width=0.33\textwidth]{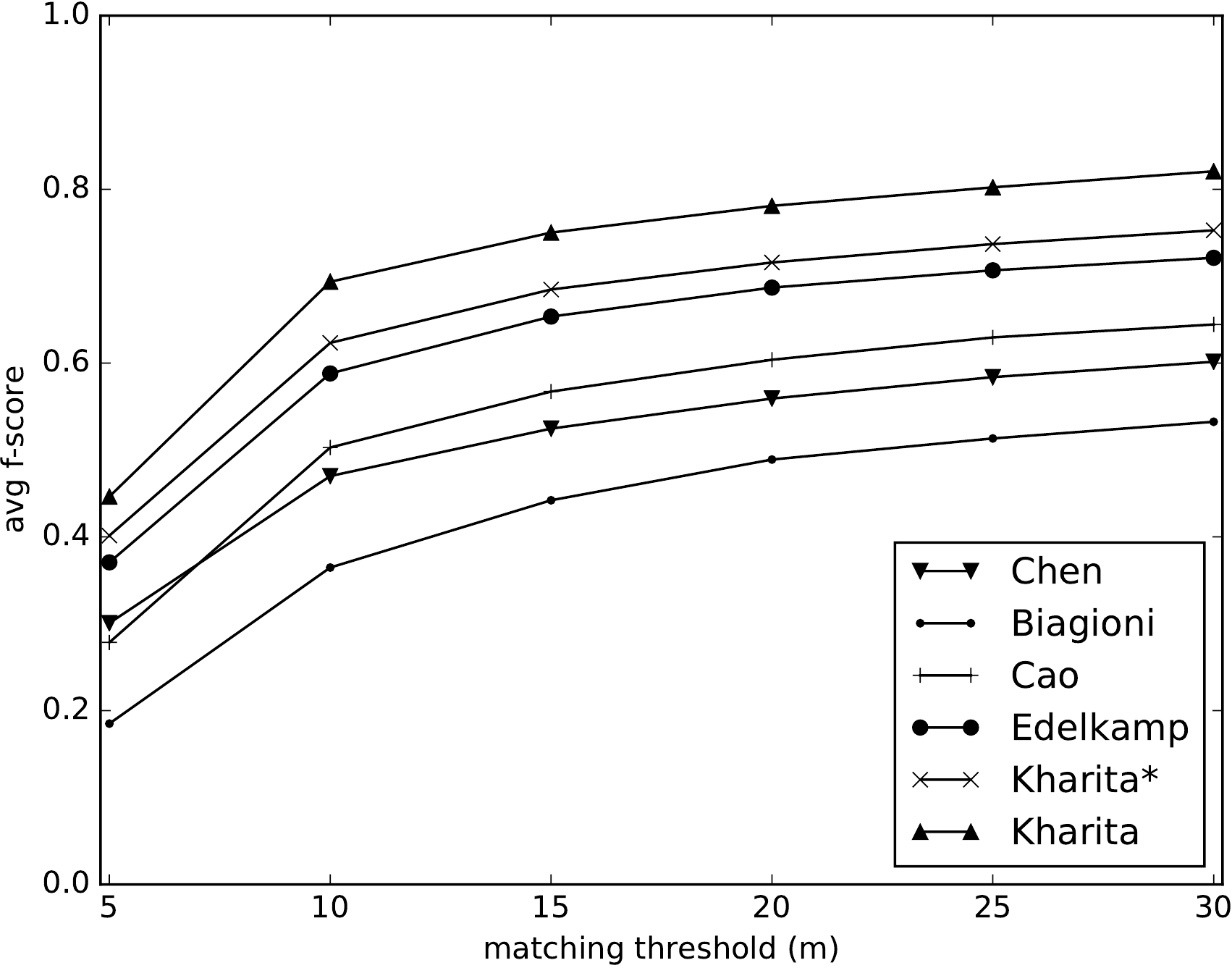}
  \includegraphics[width=0.33\textwidth]{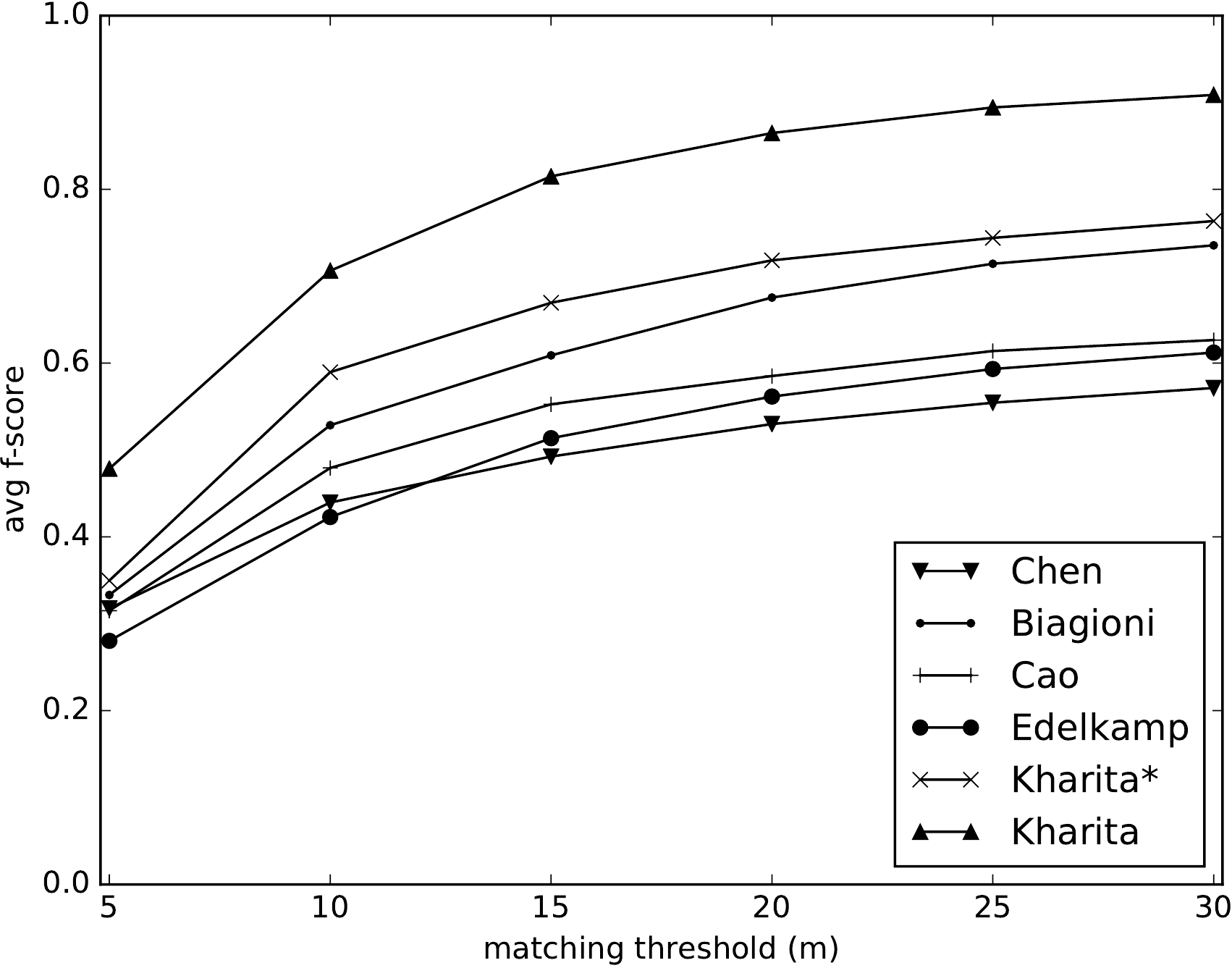}
   \caption{TOPO $f_{score}$ for different map-inference solutions. (top) Doha (1-day), (bottom) UIC data}
   \label{fig:TOPO_SOTA}      
\end{figure}

The true test of the map quality is the TOPO method. It measures how accurately intersection are inferred with the corresponding connectivity among different road segments. In Figure \ref{fig:TOPO_SOTA} we report the TOPO $f_{score}$ for the same six maps and both datasets. We observe that \kha and \khastar manage to achieve largest $f_{score}$ among the examined algorithms. This improvement comes from the improved accuracy of inferring connections (at the intersections) between the road segments; a task at which most of the previous methods fail, especially when oblivious of incoming trajectory information. 

\kha achieves TOPO $f_{score}$ (with matching radius of $30m$) of 0.91 on the UIC dataset and 0.8 on the Doha dataset which improves state-of-the-art by 20\% and 10\%, respectively. In general, TOPO $f_{score}$ are greater on the UIC dataset compared to Doha. 
This is the result of the fact that data is generated by buses following (mostly) regular routes, hence the network structure is fairly regular with many trajectories covering most inferred road-segments. In contrast, Doha dataset is much more diverse both in terms of routes and intersection types, and hence more difficult to accurately infer. Another interesting observation is that despite the fact that \khastar is presented with a very limited future information regarding the incoming trajectories, due to the online streaming constraints, it was able to achieve higher $f_{score}$ (up to 0.75) than all other algorithms except \kha which acts offline. Finally, note that some solutions perform well on one of the datasets and not so well on the other suggesting that their approach is data-dependent, while \kha effectively handles both datasets.
\begin{figure*}%[ht]
\centering
\includegraphics[width=\linewidth]{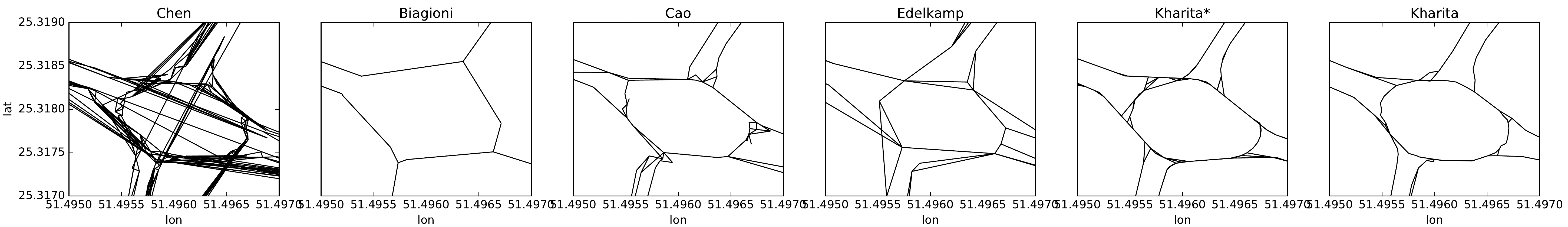}
\vspace{-3mm} 
\caption{The six maps at prominent TV roundabout in Doha.}
\vspace{-3mm} 
\label{fig:tvroundabout_4maps}
\end{figure*} 

\subsubsection{Visual Comparison}

We depict the output of the 6 mapping algorithms on the same roundabout in Figure \ref{fig:tvroundabout_4maps}. Note that each map has its own way of creating edges which determines the final quality of the map. Both \kha and \khastar faithfully map the underlying road geometry and connectivity, with very few redundant edges. A visual inspection of the roundabout generated by Biagioni algorithm explains its poor GEO $f_{score}$ performance as it misses too many road segments. Likewise, the fact that Chen's algorithm generates too many undesirable edges increases its GEO $f_{score}$. However, because the edges are not correctly connected, its TOPO $f_{score}$ performance is not very good. It is important to notice that for navigation purposes, the most important aspect of the map to be captured is the connectivity, which translates in the case of complex intersections and roundabouts to being able to correctly infer all in/out links and turns.

\subsubsection{Time performance of SOTA algorithms}
Another important aspect to explore is the time performance of the different solutions. Table~\ref{tab:timePerformance} reports the execution time of the six algorithms to process 1 days of data from Doha and one month of data from UIC. Note that the two datasets are very similar in terms of number of GPS points ($\approx 200K$.) However, the rate at which points are generated is different, which is reflected on the times achieved. Doha data being denser, it requires generally more time to process compared to the sparse UIC data. 

As previously reported in \cite{biagioni2012inferring}\cite{ChenLHYGG16}, Edelkamp, Biogioni and Cao are expensive in terms of execution time. For Edelkamp, the algorithm runs 10 iterations in the case of Doha data and 19 iterations in the case of UIC data before it converges. The time we report in Table~\ref{tab:timePerformance} as well as the $f_{score}$ reported previously are all the results of these iterations. Cao needs almost one hour in the case of Doha to make one iteration of its clarification step. Thus, it was unrealistic to aim for a larger number of iterations. Due to its myriad steps, Biagioni algorithm required 01h20min to digest one day of Doha data. Surprisingly, our two solutions \kha and \khastar outperformed all existing algorithms. The reason for which \kha runs in less time than its online counterpart \khastar is due to the use of the efficient ``bash'' ball-search in the KD-tree index to find nearest neighbors, not possible in the online setting as the data comes one point at a time.  

Next, we will explore how GEO and TOPO $f_{score}$ scale with additional data for \kha (such study is not possible for other algorithms, due to scalability issues mentioned above).

\begin{table}[ht]
\small
\caption{Comparative results for time performance}
\begin{center}
\begin{tabular}{|c|c|c|}
\hline
\textbf{Algorithm} & Doha (1 day) & UIC (1 month) \\ 
\hline
Edelkamp & 6,621$^sec$ & 1,254$sec$ \\
\hline
Cao & 3,558$sec$ & 1,637$sec$ \\
\hline
Biagioni & 4,719$sec$ & 1,280$sec$ \\
\hline
Chen & 1,078$sec$  & 381$sec$   \\
\hline
\kha & 167$sec$  & 73$sec$  \\
\hline
\khastar & 217$sec$ & 64$sec$ \\
\hline
\end{tabular}
\end{center}
\label{tab:timePerformance}
\end{table}

\subsection{\kha at scale}

In previous paragraphs we compared the \kha against other map inference solutions. In this section we will examine how the quality of the maps changes when more data is available and look into computational scalability of our algorithms. 

We run both \kha and \khastar on slices of 1 day, 7 days, and 30 days of Doha data. We report for each slice the amount of GPS points processed, the execution time, and both GEO and TOPO $f_{scores}$.

We empirically observe that the execution times scale close to linearly with the size of the input dataset making the proposed algorithms highly scalable. 

In terms of $f_{score}$, with more data GEO $f_{score}$ improves for both \kha and \khastar as simply more data implies better coverage of the map. More interestingly, we observe that \kha TOPO $f_{score}$ also improves implying \kha's ability to improve the topological accuracy when additional data is available. Oddly, \khastar TOPO performance suffers with more data due to the creation of spurios edges. Since \khastar is online in nature, it is designed to work over short time windows and poor scalability is not a major concern of \khastar.

\begin{table}[ht]
\small
\caption{\kha at scale}
\begin{center}
\begin{tabular}{|c|c|c|c|}
\hline
\textbf{$\rightarrow$ \# days} & 1 & 7 & 30 \\ 
\hline
 \multicolumn{3}{l}{Data}\\
\hline
\# GPS points & 195,283  & 1,295,360  & 5,570,806  \\
\hline
\# Trajectories & 2,834  & 22,907  & 77,314  \\
\hline
 \multicolumn{3}{l}{Time performance (seconds)}\\
\hline
\kha & 167  & 1,543  & 8,190   \\
\hline
\khastar & 217 & 1,798 & 5,512  \\
\hline
 \multicolumn{3}{l}{GEO F1 scores - 30$m$ matching radius}\\
\hline
\kha & 0.63  & 0.72  & 0.8 \\
\hline
\khastar & 0.63 & 0.73 & 0.8 \\
\hline
 \multicolumn{3}{l}{TOPO F1 scores - 30$m$ matching radius}\\
\hline
\kha & 0.8 & 0.85 & 0.86\\
\hline
\khastar & 0.76 & 0.74 & 0.73 \\
\hline
\end{tabular}
\end{center}
\label{tab:KharitaAtScale}
\end{table}

\section{Related Work}
\label{sec:relatedWork}

Due to the widespread availability of smartphones and the advent of autonomous cars, 
the problem of map construction from opportunistically collected GPS traces have been extensively study by 
various communities including data mining, geo spatial computing, transportation and computational geometry
\cite{edelkamp2003route,davies2006scalable,cao2009gps,BiagioniE12,biagioni2012inferring,ChenLHYGG16,ahmed2012constructing}.
In this section, we provide a representative summary while additional details can be found in surveys such as 
\cite{biagioni2012inferring, ahmed2015comparison, LiuBEWFZ12}.

Most prior work on map construction can be divided into three categories \cite{biagioni2012inferring}.
K-Means based algorithms perform clustering over the GPS points (typically, latitude and longitude, but sometimes also the direction).
Once the clustering converges, the centroids are linked to get a routable map.
Representative algorithms include \cite{edelkamp2003route,agamennoni2011robust,schroedl2004mining}.
Kernel density estimation (KDE) based algorithms such as \cite{chen2008roads,davies2006scalable,shi2009automatic} 
transform the input GPS points into a density discretized image that is then used to construct maps through image processing algorithms such as centerline detection.
Finally, trace merging based approaches such as \cite{cao2009gps,ahmed2012constructing}
start with an empty map and incrementally insert traces into it based on distance and direction.
Our offline algorithm \kha is a hybrid algorithm that combines $k$-Means clustering followed by trajectory processing similar to trace merging.
\cite{BiagioniE12} proposed a hybrid pipeline based on KDE approach along with adaptive thresholds, geometry and topology refinement.
\cite{ChenLHYGG16} proposed a supervised learning framework that can leverage prior knowledge on real-world road networks.
There has been a series of papers that can infer additional road metadata such as intersections, number of lanes, speed limit, road type etc
\cite{schroedl2004mining,niehoefer2009gps}.

There has been handful of work \cite{schroedl2004mining, zhang2010integration, wang2013crowdatlas,bruntrup2005incremental,ahmed2012constructing,vanall}
on maintaining and updating maps when new GPS data points arrive.
However, most of these approaches do not have good practical performance 
and are very sensitive to differential sampling rates, disparity in data points, GPS errors etc.
Often, these algorithms seek to directly extend one of the three approaches and 
suffer from bottlenecks arising from algorithmic step that is fundamental to it (such as clustering, density estimation, clarification, map matching) etc.
As illustrated by Table~\ref{tab:timePerformance} that summarized the comparative performance of the algorithms, 
simple adaptations of those algorithms cannot update/refine maps in a near real-time manner. 
In contrast, our online algorithm \khastar uses recent advances in online algorithms 
for $k$-Means \cite{DBLP:conf/alenex/2016} and graph spanners \cite{McGregor:2014} to circumvent the performance bottlenecks.

Algorithms for spanners have been developed for general graphs\cite{peleg1989graph}, geometric graphs \cite{narasimhan2007geometric} 
and in streaming settings\cite{McGregor:2014}.
The distance approximating functionality of spanners has multiple applications including 
robotics motion planning\cite{wang2015fast}, telecommunication network design, 
serving as distance oracle for proximity problems\cite{narasimhan2007geometric,peleg1989graph} etc.
However, our work is the first to introduce graph spanners in the context of map inference.

\section{Conclusion}
\label{sec:conclusion}

In this paper, we proposed two efficient algorithms \kha and \khastar for constructing maps from opportunistically collected GPS information. The first is a two-phase algorithm that clusters GPS points followed by a sparse graph construction using spanners. The second is an online algorithm that can create and update the map when the GPS data points arrive in a streaming fashion. While our approach is conceptually simple it significantly outperforms the state-of-the-art due to efficient exploitation of angle and speed information and elegant handling of geographic information (via clustering) and topological structure (via graph spanner). 

\vspace{-1mm} 
{%\tiny 
\bibliographystyle{ACM-Reference-Format}
\bibliography{mapinf} 
}
\end{document}